\documentclass[10pt,twocolumn]{article}

\usepackage[utf8]{inputenc}
\usepackage[english]{babel}
\usepackage[T1]{fontenc}
\usepackage{amsmath,amsfonts,amssymb}
\usepackage{mathpazo} 
\usepackage[scaled]{helvet}
\usepackage{graphicx,xcolor}
\usepackage{booktabs}
\usepackage{siunitx}
\usepackage[colorlinks=true, allcolors=blue]{hyperref}

\usepackage[numbers,sort&compress]{natbib}
\usepackage{doi}
\usepackage{geometry}
\usepackage{authblk}
\usepackage{subcaption}
\usepackage{glossaries}
\usepackage[numbers,sort&compress]{natbib}
\definecolor{affil}{RGB}{0,0,0} 
\definecolor{author}{RGB}{59,90,198}

\setacronymstyle{long-short}
\newacronym{pl}{PL}{photonic lantern}
\newacronym{mspl}{MSPL}{mode-selective photonic lantern}
\newacronym{itr}{ITR}{inverse taper ratio}
\newacronym{lp}{LP}{linearly polarized}
\newacronym{dcf}{DCF}{double-clad fiber}
\newacronym{na}{NA}{numerical aperture}
\newacronym{oam}{OAM}{orbital–angular-momentum}
\newacronym{pm}{PM}{polarization-maintaining}

\begin{document}
\title{\sffamily\bfseries Simultaneous plane illumination and detection in confocal microscopy using a mode-selective photonic lantern}
\author[1,2,3,*]{Rodrigo Itzamn\'a Becerra-Deana}
\author[1,*]{Simon Desrochers}
\author[1]{Rapha\"el Maltais-Tariant}
\author[4]{Simon Brais-Brunet}
\author[1,3]{Guillaume Ramadier}
\author[1]{St\'ephane Virally}
\author[1,3]{Lucien E. Weiss}
\author[1,2,3,4,**]{Caroline Boudoux}

\affil[1]{Polytechnique Montr\'eal, 2500 Chemin de Polytechnique, Montr\'eal, QC H3T 1J4, Canada}
\affil[2]{Castor Optics, 361 Boulevard Montpellier, Saint-Laurent, QC H4N 2G6, Canada}
\affil[3]{Peregrine Photon, Montr\'eal, Canada}
\affil[4]{Université de Montr\'eal, Institute of Biomedical Engineering, Montr\'eal, QC H3T 1J4, Canada}
\affil[*]{The authors contributed equally to this work.}
\affil[**]{caroline.boudoux@polymtl.ca}

\date{\today}

\maketitle

\begin{abstract}
Confocal microscopy is the cornerstone of cellular biology and biomedical research due to its non-destructive imaging, compatibility with live cells, sensitivity, optical sectioning, and subcellular resolution. To meet the demand for rapid three-dimensional imaging, we propose a novel approach using a \gls{mspl}. This fiber-based device transforms single-mode light into multiple linearly polarized modes, allowing simultaneous detection of multiple planes. Using a four-port \gls{mspl} to manipulate three group modes (LP$_{01}$, LP$_{11}$, and LP$_{21}$), we demonstrate high-throughput imaging simultaneously with multiple planes. This technique exploits differences in focus sections across modes, enabling individual multi-plane detection via a spatial division multiplexer, with some trade-off in resolution and field of view.
\end{abstract}

\section{Introduction}

Confocal microscopy is a powerful imaging technique that offers non-destructive analysis, high spatial resolution, and versatility. It is used for biological and material samples across diverse scientific disciplines. These features make confocal microscopy indispensable in cellular biology, material science, and biomedical research, as well as in generating high-resolution surface topographies \cite{nwaneshiudu_introduction_2012, wright_chapter_2002, bayguinov_modern_2018, st_croix_confocal_2005, newton_optimisation_2023, udupa_characterization_2000, bezak_identification_2014, jordan_highly_1998, netuschil_pilot_1998}

However, the growing demand for rapid 3D acquisition and high-resolution imaging has generated significant innovation in optical microscopy. One prominent strategy is the use of simultaneous detection techniques—such as multiple pinholes and beam splitters—to capture several optical planes at once. This parallelized approach increases data throughput and minimizes imaging time, though it can suffer from reduced photon detection when employing more than two beam splitters \cite{zhao_high-resolution_2023, tsang_fast_2021, weber_high-speed_2023, weber_multi-plane_2022, wang_non-axial-scanning_2017}. Another parallelized method is chromatic confocal microscopy, which exploits the chromatic aberration of a lens to focus simultaneously at different planes within a sample \cite{blateyron_chromatic_2011, shi_chromatic_2004, li_dmd-based_2020, jeong_swept-source-based_2020, sharma_ultralong_2023}. Additional multiplane confocal imaging strategies include the use of acoustic tunable lenses \cite{duocastella_simultaneous_2014, duocastella_simultaneous_2015}, z-splitter prisms \cite{xiao_high-contrast_2020}, multiplexed volume holographic gratings \cite{chia_multiplexed_2018}, and quadratically distorted gratings \cite{dalgarno_multiplane_2010}.

In this study, we introduce a novel strategy that enables simultaneous illumination and detection of multiple planes by leveraging higher-order optical fiber modes. In our approach, each z-position is encoded at a unique spatial location in the detection path using a fiber-based spatial division multiplexer, known as a \gls{mspl}. \Glspl{mspl} are innovative fiber-based (de)multiplexers capable of efficiently transforming single-mode light into specific linearly polarized (LP) modes within a few-mode fiber section \cite{birks_photonic_2015, leon-saval_mode-selective_2014, simuCUI2023129550, becerra-deana_mode-selective_2024}. This modal transformation achieves high efficiency, with losses below 10\% over a broadband spectral range of 500~\unit{nm}~\cite{becerra-deana_mode-selective_2024, fontaine_photonic_2022, 3PL_Becerra-Deana:25}, making \glspl{mspl} particularly beneficial for biomedical applications such as optical coherence tomography~\cite{sivry-houle_all-fiber_2021, Maltais-Tariant:23, Maltais-Tariant:25}. \Glspl{mspl} generate multiple co-aligned beams optimized for simultaneous illumination and detection. By exploiting the intrinsic differences between modes—such as spatial distribution and \gls{na}—we can distinguish among planes. This approach accelerates scanning acquisition compared to conventional confocal microscopy but presumnably still slower than widefield scanning, by requiering a single x-y scan to capture multiple z planes in parallel using spatial-division multiplexing. Employing a four-port \gls{mspl} allows manipulation of three distinct group modes (LP01, LP11, and LP21), enabling concurrent acquisition of several planes, while accepting inherent trade-offs in resolution and field of view. Moreover, this technique is extensible beyond reflective confocal microscopy: the implementation of a visible photonic lantern could readily extend its impact to fluorescence imaging without substantial complications.

\section{Photonic Lantern and Confocal Microscopy}

\begin{figure}[ht]
\centering
\includegraphics[width=\linewidth]{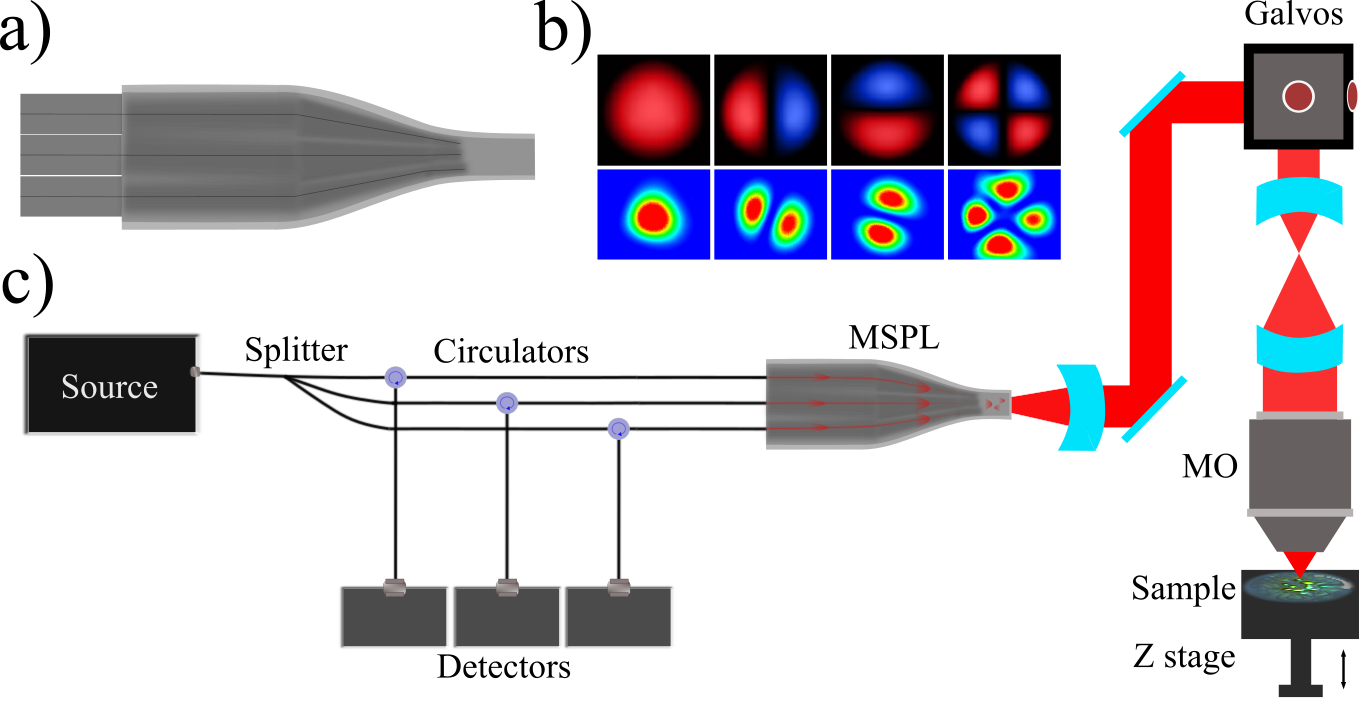}
\caption{Photonic lantern and setup schematics. a) Side view of the fused-tapered fiber bundle, showing the four generated modes. b) A confocal system that implements a \gls{mspl}, featuring a splitter designed to illuminate all ports simultaneously, along with three circulators to independently detect each plane.}
\label{fig:SetupMP}
\end{figure}

The \gls{mspl} was fabricated and characterized using four custom double-clad fibers, as in \cite{becerra-deana_mode-selective_2024}, one extra fiber that the three presented in that paper. This \gls{mspl} exhibits an excess loss of less than -0.5~\unit{dB} and a high modal isolation of at least 20~\unit{dB} per group mode. Figure~\ref{fig:SetupMP}a) presents a schematic of a \gls{mspl}, consisting of several optical fibers arranged within a capillary tube that is tapered and fused, becoming the new few-mode waveguide. Figure~\ref{fig:SetupMP}b) shows the modes that this \gls{mspl} can generate. At the top, the simulated LP modes are displayed, while the experimental far-field images of the modes are shown at the bottom. Figures~\ref{fig:SetupMP}c) shows the confocal system used with the three group modes of the \gls{mspl}. It consists of a laser source (Thorlabs S5FC1021S, USA) connected to a splitter (JDS Uniphase ACWA104, USA) that directs light to three different circulators connected to the ports of the \gls{mspl}. The output from the lantern is collimated and directed through galvanometric mirrors, then passes through a telescope that matches the back focal plane of the microscope objective (Nikon CFI Plan Apo 20x/0.75 DIC, Japan) to the LP$_{21}$ mode, and finally reaches the sample. Detection is performed through the same optical path as the \gls{mspl} demultiplexes the signal to each fiber port, connected to the circulators, and then sent to the photodetectors (Thorlabs PDA10CS, USA) and (Thorlabs PDA20CS2, USA). 

\section{Plane shift}

Each mode exhibits a distinct beam width that increases with mode number, a consequence of differences in effective refractive index at the output of the \gls{mspl}. This leads to unique spatial distributions and \glspl{na} for each mode. In the system, the LP$_{21}$ mode was chosen to match the back focal plane of the microscope objective because the LP$_{01}$ mode produces a narrower beam waist. Measurements with the knife-edge method indicate that the beam width of LP$_{21}$ is 1.05 times larger than that of LP$_{01}$.

This variation in \glspl{na} can be leveraged to shift the focal positions of beams from different group modes. Table~\ref{tab:beamwaist} presents simulations using Gaussian beam wave optics~\cite{lodewyck_gaussianbeam_2013} for a two-lens optical system, where small differences in initial beam \glspl{na} result in distinct waist positions at the sample. The simulations were conducted at a wavelength of 1300~\unit{nm}, with collimating and imaging lenses of 4.7~\unit{mm} and 5.7~\unit{mm} focal lengths, respectively. The collimating lens was placed 4.7~\unit{mm} after the initial waist, and the imaging lens was located 200~\unit{mm} beyond the collimating lens. These results demonstrate how minor variations in initial divergence can translate into measurable shifts in the focal plane.

\begin{table}[ht]
    \centering
    \begin{tabular}{cc|c}
         Initial beam & &Beam at sample  \\
         Waist size & Divergence & Waist Position \\
         \hline
         6~\textmu m &69.4~mrad &210.416~mm \\
         5.7~\textmu m &73.0~mrad &210.456~mm \\
    \end{tabular}
    \caption{Simulation of the same two-lens optical system with a slight variation in the initial beam \gls{na} and its impact on the waist position at the system's exit.}
    \label{tab:beamwaist}
\end{table}

\begin{figure}[ht]
\centering
\includegraphics[width=\linewidth]{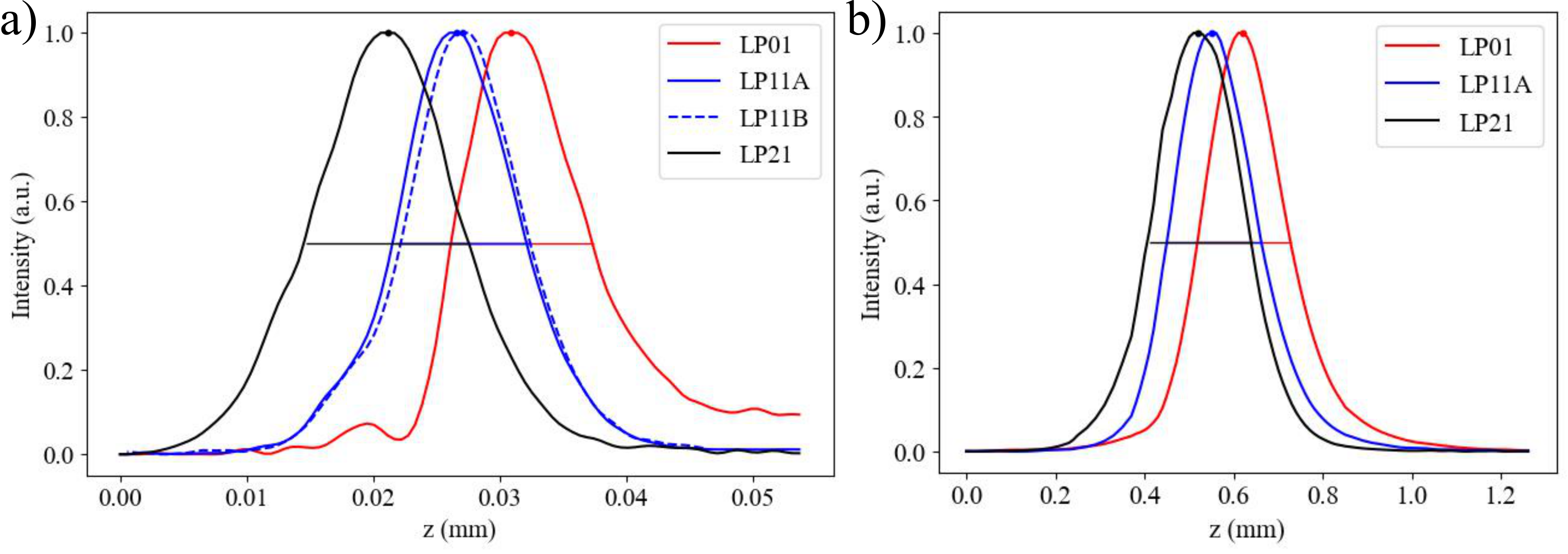}
\caption{Axial resolution using a) 20X and b) 3X microscope objectives for each mode of a \gls{mspl}. Each plot represents the LP mode LP$_{01}$ as solid red, LP$_{11a}$ as dashed blue, LP$_{11b}$ as solid blue, and LP$_{21}$ as solid black.}
\label{fig:AxialResolution}
\end{figure}

To experimentally confirm focal plane shifts, we implemented the setup depicted in Figure~\ref{fig:SetupMP}b, placing a mirror at the sample position and performing a z-scan with a motorized stage. Figure~\ref{fig:AxialResolution}a shows the detected power at each plane, measured by photodetectors when all ports are simultaneously illuminated. The data are grouped by optical mode: LP$_{01}$ (solid red), LP$_{11a}$ (dashed blue), LP$_{11b}$ (solid blue), and LP$_{21}$ (solid black), using a 20X microscope objective. The y-axis represents intensity (arbitrary units), while the x-axis indicates position (\unit{mm}). For comparison, Figure~\ref{fig:AxialResolution}b presents the results with a 3X objective, revealing a similar trend. All point-spread functions (PSFs) were smoothed with a Gaussian filter and normalized to their respective amplitude ranges.

With the 20X microscope objective, the optical shift between LP$_{01}$ and LP$_{11a}$ was measured to be 4.3~\unit{\micro m}, while the shift between LP$_{11a}$ and LP$_{21}$ was 5.4~\unit{\micro m}. In contrast, using the 3X objective, the shift between LP$_{01}$ and LP$_{11a}$ increased to 70.1~\unit{\micro m}, and the shift between LP$_{11a}$ and LP$_{21}$ was 30~\unit{\micro m}. These differences arise because each mode propagates differently through the optical system, due to variations in spatial distribution and propagation constants, which impact their interaction with different optical components, especially the microscope objective.

As shown in Figure~\ref{fig:AxialResolution}, no significant focal plane shift is observed between the degenerate LP$_{11}$ modes. This result is expected, given their nearly identical \gls{na} and propagation constants, which lead to negligible differences in focal plane position. Therefore, our analysis focuses on the three distinct groups of modes where clear separation occurs.

\begin{figure}[ht]
\centering
\includegraphics[width=\linewidth]{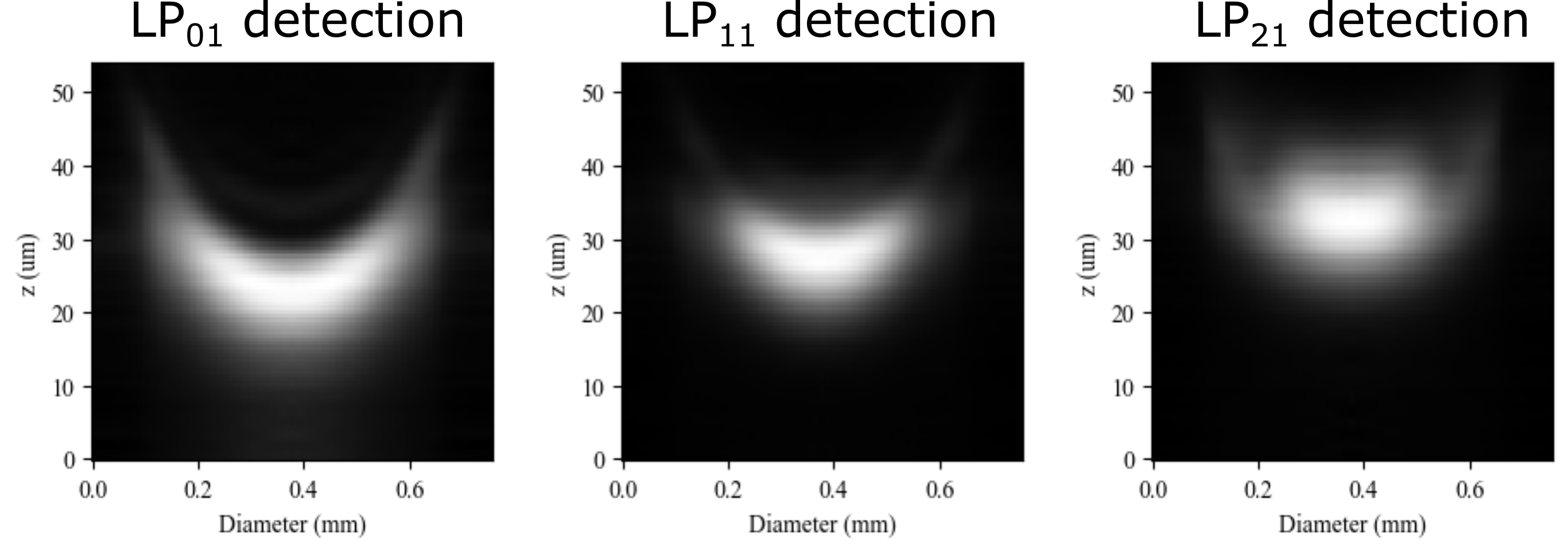}
\caption{Miror scanned confocal system. Detection performed with LP$_{01}$, LP$_{11}$, and LP$_{21}$,}
\label{fig:scaning}
\end{figure}

Figure~\ref{fig:scaning} presents detection results while scanning in the XZ plane for (a) LP$_{01}$, (b) LP$_{11}$, and (c) LP$_{21}$ modes using galvanometric mirrors with the 20X microscope objective. This approach enables determination of the field of view, which is reduced to 74\% compared to LP$_{01}$. The calculation is based on the full width at half maximum (FWHM) across the x-axis at the brightest point, comparing the smallest and largest FWHM (with LP$_{01}$ providing the largest). Additionally, the power distribution along the Z axis for LP$_{21}$ reveals a blurrier region, which is consistent with findings in ~\cite{SarahMSPLConfocal}, indicating that a higher-order mode experience a 9\% lateral and 3\% axial resolution loss—limitations that increase with mode number.

\section{Multiplane imaging with an mspl}

\begin{figure}[ht]
\centering
\includegraphics[width=\linewidth]{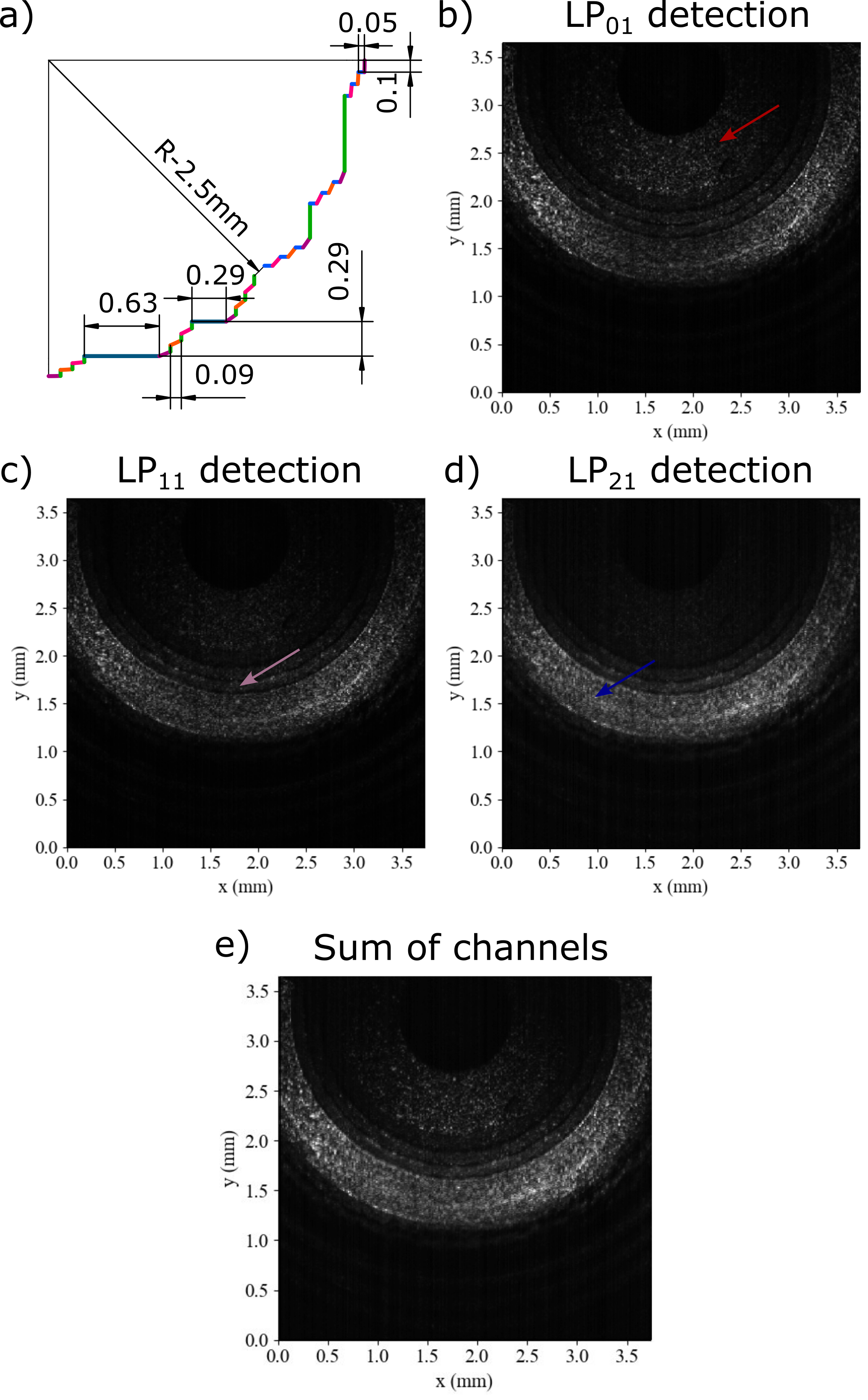}
\caption{Multiplane sample. a) shows a diagram of the custom sample. Measurements are in millimeters.  Simultaneous detection of each plane is exhibited by a) LP$_{01}$, b) LP$_{11}$, and c) LP$_{21}$. d) Maximum intensity projection of the three detceted images.}
\label{fig:Multiplane}
\end{figure}

To experimentally demonstrate the multiplane illumination and detection scheme, we used a nylon custom-made calibration target defined in previous work \cite{brais-brunet2025a}. This hemispherical target was designed and fabricated with well-defined steps and side sections. Figure~\ref{fig:Multiplane}a presents a schematic of the sample, with all dimensions provided in millimeters. In this work, the lower part of the target was imaged, with the two largest step being the object of focus. Using the 3X microscope objective, we simultaneously illuminated and detected multiple planes within the sample. Figures~\ref{fig:Multiplane}b, c, and d display the images corresponding to the LP$_{01}$, LP$_{11}$, and LP$_{21}$ modes, each capturing a different focal plane and structural detail. Finally, Figure~\ref{fig:Multiplane}e shows a maximum intensity projection, combining information from all three detected planes into a single comprehensive image.

A key advantage of this method is that the demultiplexing process enables both illumination and collection from specific z positions in a single output easily. In contrast, conventional multiplane detection splitting is disconnected from the illumination, leading to additional sample exposure and significantly increasing the complexity of the setup required to align illumination and collection perfectly. Figures~\ref{fig:Multiplane}b, c, and d clearly show that each mode provides unique contributions: LP$_{01}$ predominantly images deeper planes (as indicated by the red arrow in Figure~\ref{fig:Multiplane}b), with high intensity in the lower steps and minimal response in higher planes such as those captured by LP$_{21}$. Conversely, LP$_{21}$ excels at imaging higher planes (blue arrow in Figure~\ref{fig:Multiplane}d), with enhanced intensity in the upper steps compared to LP$_{01}$. The LP$_{11}$ mode captures both lower and upper structures, providing a more balanced view of intermediate depths, although its intensity is not sharply peaked in either region or in the middle region (pink arrow in Figure~\ref{fig:Multiplane}c). Notably, the structural features are well-resolved in the LP$_{01}$ and LP$_{11}$ images, while the LP$_{21}$ image appears more blurred—reflecting the reduced resolution associated with higher-order modes. This loss in resolution could potentially be mitigated through computational post-processing techniques, such as PSF engineering by further investigating higher-order modes illumination and detection. 

Despite the presence of significant speckle contrast in the images, this effect can be minimized by employing a visible \gls{mspl} for fluorescence imaging in biological applications or by using a broader bandwidth light source for material analysis, as \gls{mspl} has a broadband response. Importantly, the maximum intensity projection (Figure~\ref{fig:Multiplane}e) effectively combines information from all three planes, offering a comprehensive view of the sample with a single XY scan. This approach reduces acquisition time by a factor equal to the number of modes used (three in this case), and scaling up to more modes could further accelerate imaging speed—an important advantage for high-throughput applications.

\section{Conclusion}

In this work, we have demonstrated a novel approach to simultaneous multiplane imaging in confocal microscopy using a \gls{mspl}. This \gls{mspl}, characterized by high modal isolation and low excess loss, enables efficient demultiplexing of three distinct group modes (LP${01}$, LP${11}$, and LP$_{21}$), each corresponding to a different imaging depth within the sample. By exploiting the spatial distribution and \gls{na} of the modes, we have shown that focal position differences among LP modes in the few-mode output can be harnessed for concurrent illumination and detection across multiple planes. This results in a significant reduction in acquisition time, which scales with the number of modes used.

Through a combination of simulation and experimental validation, we demonstrated that small differences in the \gls{na} of each mode can be used to achieve controlled focal plane separation, with the magnitude of this separation depending on the microscope objective. Multiplane imaging experiments on a structured custom sample confirmed the capability of this technique to resolve different depths simultaneously, with each mode contributing distinct plane information.

The principal limitations of this approach are the reduction of field of view and resolution for higher-order modes, as well as some resolution loss from incomplete back focal plane filling with LP$_{01}$. However, these challenges can be mitigated through computational post-processing, optimized sample selection, or restricting analysis to regions of interest. The method is also highly versatile, with strong potential for extension to fluorescence imaging and broadband sources. This is particularly impactful for fluorescence microscopy—one of the most widely used techniques for biological samples—and for multichannel applications enabled by the broadband response of \glspl{mspl}.

Regarding scalability of the modes, \glspl{mspl} have been shown to support up to five group modes~\cite{velazquez-benitez_scaling_2018}. Alternative demultiplexing techniques, such as spatial light modulators or multiplane light conversion, may offer more straightforward scalability at the cost of efficiency, providing flexibility for different experimental requirements.

Overall, our results establish \gls{mspl}-enabled multiplane confocal microscopy as a powerful technique for high-throughput, depth-resolved imaging. This advance opens new possibilities for rapid volumetric imaging in biological, biomedical, and materials science applications, paving the way for future developments in multimodal and high-content microscopy.

\newcommand{\backmatter}{
    \bigskip
    \section*{Backmatter} 

\section*{Funding} CB and SV acknowledge funding from the Mid-Infrared Quantum Technology for Sensing (MIRAQLS) project, supported by the European Union’s Horizon Europe research and innovation programme under grant agreement 101070700. Natural Sciences and Engineering Research Council (NSERC) of Canada grants \#RGPIN-2018-06151 (CB), I2IPJ 590717 - 24 (LEW and CB). LEW acknowledges a salary award from the Fonds de Recherche du Québec– Santé (FRQS).

\section*{Disclosures} RIBD, GR, LEW, and CB: Peregrine Photon Inc. (I), and Polytechnique Montreal (P). RIBD: Castor Optics Inc. (E) CB: Castor Optics Inc. (I)

\section*{Data Availability Statement}  Data underlying the results presented in this paper is not publicly available at this time but may be obtained from the authors upon reasonable request.

}

\bibliography{Multiplane_confocal, Confocal_review, Topography_confocal}

\end{document}